\documentclass{svjour2}%
\usepackage{graphicx}
\usepackage{amsmath}
\usepackage{amsfonts}
\usepackage{amssymb}%
\smartqed
\begin{document}
%
\title
{Quantum decoherence in a pragmatist view: Resolving the measurement problem
}%
%

\titlerunning{Resolving the measurement problem}%
%

\author{Richard Healey}%
%

\authorrunning{Richard Healey}%
%

\institute
{Philosophy Department, University of Arizona, Tucson, Arizona 85721-0027, USA \\
Tel.: +520-621-6045\\
Fax: +520-621-9559\\
\email{rhealey@email.arizona.edu}
}%
%

\date{Received: date / Accepted: date}%
%

\maketitle
%

\begin{abstract}
This paper aims to show how adoption of a pragmatist interpretation permits a
satisfactory resolution of the quantum measurement problem. The classic
measurement problem dissolves once one recognizes that it is not the function
of the quantum state to describe or represent the behavior of a quantum
system. The residual problem of when, and to what, to apply the Born Rule may
then be resolved by judicious appeal to decoherence. This can give sense to
talk of measurements of photons and other particles even though quantum field
theory does not describe particles.%

\end{abstract}%


\section{Introduction}

\label{intro}

Attitudes to the quantum measurement problem vary widely. Optimists deny there
is any problem, while pessimists maintain there is a serious problem that
could be solved only by modifying quantum theory. An intermediate opinion is
that while quantum theory does face a \textit{prima facie} measurement
problem, this problem is readily resolved by applying the unmodified theory to
interactions that efficiently delocalize the phase of a system into its
environment during a quantum measurement. Though widespread, this opinion
remains controversial. Attempts to support it by argument have met objections
from both optimists and pessimists intended to show why simply applying
quantum theory to establish the rapid and effectively irreversible
diagonalization of the reduced density operator of a system does not solve any
problem posed by measurement in quantum theory.

This paper aims to show how adoption of a pragmatist interpretation permits a
satisfactory resolution of the quantum measurement problem. The classic
measurement problem dissolves once one recognizes that it is not the function
of the quantum state to describe or represent the behavior of a quantum
system. The residual problem of when, and to what, to apply the Born Rule may
then be resolved by judicious appeal to decoherence. This can give sense to
talk of measurements of photons and other particles even though quantum field
theory does not describe particles.

\section{The classic measurement problem\label{section 2}}

The quantum measurement problem is to reconcile quantum theory with the fact
that measurements have outcomes displaying patterns the theory is supposed to
predict and (as far as possible) explain. The problem is most acute if one
assumes that a system's quantum state gives a complete description of that
system, for then the linearity of quantum theory implies that measurements
typically fail to have outcomes, in manifest contradiction to experimental reports.

Discussions of measurement in quantum theory often begin with an idealized
model discussed by von Neumann\cite{1} in which a quantum state of a system
$S$ becomes correlated with that of a (quantum) apparatus system $A$. In
a\ simple 2 qubit version of this model $S$ acts as a \textit{c}-NOT gate as
follows:
\begin{align}
\left\vert +\right\rangle _{S}\otimes\left\vert \Uparrow\right\rangle _{A}  &
\rightarrow\left\vert +\right\rangle _{S}\otimes\left\vert \Uparrow
\right\rangle _{A}\label{1}\\
\left\vert -\right\rangle _{S}\otimes\left\vert \Uparrow\right\rangle _{A}  &
\rightarrow\left\vert -\right\rangle _{S}\otimes\left\vert \Downarrow
\right\rangle _{A}.\nonumber
\end{align}
By linearity, its action on a non-trivial superposed state ($\left\vert
a\right\vert ^{2},\left\vert b\right\vert ^{2}\neq0,1$) of $S$\ is then%
\begin{equation}
\left(  a\left\vert +\right\rangle +b\left\vert -\right\rangle \right)
\otimes\left\vert \Uparrow\right\rangle \rightarrow a\left(  \left\vert
+\right\rangle \otimes\left\vert \Uparrow\right\rangle \right)  +b\left(
\left\vert -\right\rangle \otimes\left\vert \Downarrow\right\rangle \right)
\label{2}%
\end{equation}
where the system subscripts have been omitted. According to the completeness
assumption, equations (\ref{1}) describe an interaction between an apparatus
in some definite condition (say $\Uparrow$, with pointer pointing up) and a
system with some definite property (symbolized either by $+$ or by $-$) that
is undisturbed by the interaction, while the apparatus finally assumes a
definite condition correlated to, and so recording, that property.

Applied to (\ref{2}), however, the completeness assumption implies that after
the interaction the system has neither property $+$ nor $-$, and the apparatus
pointer points neither up nor down. If this were a faithful model of an actual
quantum measurement, then that measurement would record no outcome for any
non-trivial superposed state of $S$. Of course, the idealized model is wildly
oversimplified, and one might have held out hope that a more realistic model
of a quantum measurement could restore consistency with experimental
observations. But a long history of failed attempts to solve the measurement
problem by developing more realistic models of measurement as a linear
interaction between quantum system and quantum apparatus has effectively
removed that hope, as long as the completeness assumption remains in force.

Dirac\cite{2} and Von Neumann\cite{1} sought to preserve the completeness
assumption by dropping linearity. They took measurement to "collapse" or
reduce the quantum state stochastically onto an eigenstate of the measured
observable. In the simple model this means replacing (\ref{2}) by%
\begin{align}
\left(  a\left\vert +\right\rangle +b\left\vert -\right\rangle \right)
\otimes\left\vert \Uparrow\right\rangle  &  \rightarrow\text{either
}\left\vert +\right\rangle \otimes\left\vert \Uparrow\right\rangle \text{,
with probability }\left\vert a\right\vert ^{2}\text{ }\label{3}\\
&  \text{or }\left\vert -\right\rangle \otimes\left\vert \Downarrow
\right\rangle \text{, with probability }\left\vert b\right\vert ^{2}\nonumber
\end{align}

This does reconcile\ quantum theory with the fact that measurements have
outcomes. But it raises 'measurement' to the status of a primitive term in the
principles of quantum theory, needed to mark the suspension of the normal
continuous, linear dynamical evolution of the quantum state, and its temporary
replacement by a nonlinear, stochastic evolution. This disqualifies quantum
theory from giving any further account of the measurement process, leaving it
quite unclear under what physical circumstances this process is supposed to
occur. Moreover, many actual measurements do not leave the measured system in
an eigenstate of the measured observable (Pauli\cite{3} called these
measurements of the second kind), and a measurement of a photon typically
destroys or absorbs it.

Wigner\cite[chapter 12]{4} gave a classic restatement of the measurement
problem. He took the problem to be that of reconciling the continuous, linear
evolution of an unmeasured system's quantum state with its discontinuous,
stochastic evolution on measurement---in the simple model, this means
reconciling (\ref{2}) with (\ref{3}). Noting that (\ref{2}) establishes a
statistical correlation between results of possible observations on $A$ and
$S$, he concludes that while applying quantum theory to the measurement
interaction (as does (\ref{2}) in the simple model) does not (in general) show
how measurement puts the measured system in a definite state, it does enable
one to replace this task with the task of showing how observation of the
apparatus puts \textit{it} in a definite state.

\begin{quote}
The problem of measurement on the object is thereby transformed into the
problem of an observation on the apparatus. Clearly, further transfers can be
made by introducing a second apparatus to ascertain the state of the first,
and so on. However, the fundamental point remains unchanged and a full
description of an observation must remain impossible since the
quantum-mechanical equations of motion are causal and contain no statistical
element, whereas the measurement does.\cite[p.158]{4}
\end{quote}

A statistical element could be introduced into a quantum analysis of
measurement by representing the initial state of a complex apparatus by a
mixture of states from the eigenspace spanned by eigenvectors of the
(massively degenerate) pointer position observable, all with the same
eigenvalue. This could be regarded as the appropriate way of representing
one's ignorance of the exact microstate of the apparatus. But this doesn't
work, as Wigner shows:\footnote{Von Neumann \cite[pp.437-9]{1} already in 1932
gave a similar demonstration of the failure of this "often proposed" attempt
to reconcile the deterministic evolution of the quantum state with the
statistical results of observations. This may have been the first of many
demonstrations of "the insolubility of the quantum measurement problem"
involving successive removal of idealizations of the simple model (\ref{2}) of
measurement: these now include, for example, \cite{5},\cite{6},\cite{7}%
,\cite{8}.}

\begin{quote}
It must be concluded that \textit{measurements which leave the system
object-plus-apparatus in one of the states with a definite position of the
pointer cannot be described by the linear laws of quantum mechanics.
}(\cite[p.164]{4} italics in the original)
\end{quote}

This and other generalizations of the simple quantum model (\ref{2}) of
measurement establish the incompatibility between the linearity of quantum
mechanical evolution and the occurrence of definite measurement outcomes only
if the quantum state completely describes a system to which it is ascribed.
Here and elsewhere in this article Wigner seems to accept this completeness
assumption. But earlier indications that this cannot be his considered view
are confirmed by his subsequent answer to the question "What is the state
vector?" This makes it clear that he has abandoned the completeness assumption
after initially seeming to accept it.

\begin{quote}
...the state vector is only a shorthand expression of that part of our
information concerning the past of the system\ which is relevant for
predicting (as far as possible) the future behavior thereof. ... \textit{the
laws of quantum mechanics only furnish probability connections between results
of subsequent observations carried out on a system.} (\cite[p.166]{4} italics
in the original)
\end{quote}

Here he advocates a very different view of the role of the quantum state, not
as describing physical reality but as predicting our observations of it. This
promises to dissolve the classical measurement problem. We'll soon see how far
it is able to do so. But first note the rather artificial character of that
problem. Dirac and Von Neumann notwithstanding, immediate repetition of an
actual measurement of a quantum observable does not always, or even usually,
give the same result: it may disturb or even destroy the measured system. When
necessary, we apply linear quantum theory in studying how measurement affects
a system. Quantum measurements must be carefully designed if they are to
fufill their intended function. Their design involves application of the
linear laws of quantum theory itself, not some primitive stochastic law like
(3). The only useful function of Von Neumann's\cite{1} postulation of
stochastic collapse (his process 1) is to account for the possibility of
definite outcomes of quantum measurements. But there is a need for something
to play that role only if the quantum state completely describes the behavior
of a quantum system.

\section{How to dissolve it\label{section 3}}

If one accepts Wigner's\cite{4} view of the quantum state one can try to
reconcile (\ref{2}) with (\ref{3}) like this. (\ref{2}) represents the
evolution of information concerning the past of $S$ and $A$. Observation of
$A$ adds new information: either that $\Uparrow$ is true, or that $\Downarrow$
is true. On the basis of this new information, one should\ update the
information represented by the right-hand side of (\ref{2}) either to the
information represented by $\left\vert +\right\rangle \otimes\left\vert
\Uparrow\right\rangle $ or to the information represented by $\left\vert
-\right\rangle \otimes\left\vert \Downarrow\right\rangle $. The right-hand
side of (\ref{3}) correctly represents the state of information of one who
knows of the observation of $A$ but is ignorant of its result, while the
right-hand side of (\ref{2}) correctly represents the information concerning
the past of $S$ and $A$ prior to observation of $A$. (\ref{2}) and (\ref{3})
do not represent incompatible states of affairs, since no quantum state
involved represents any state of affairs.

On this view, the function of a quantum state is not to represent the
condition of a system to which it is ascribed, but to summarize information
gleaned from prior observations that is relevant to predicting the probability
of each possible outcome of any future observation. Wigner\cite{4} claims to
be reporting the orthodox view on measurement of the late 1920s: his view of
the quantum state was held by Heisenberg\cite{9}, Peierls\cite{10} and others
and has a number of contemporary advocates.\footnote{Friederich\cite{11} lists
12 recent papers developing versions of similar views in his second footnote.}
The view promises to dissolve the classic measurement problem. If one denies
that the evolution of a quantum state represents the changing physical
condition of the system to which it is ascribed, it is no longer problematic
that this state evolves in radically different ways according as the system is
or is not measured. The quantum state does not track changes in the system,
but in the available information relevant to predicting outcomes of future
measurements on the system. This information changes discontinuously as more
information becomes available, namely the results of relevant observations. In
the Heisenberg picture, the quantum state changes \textit{only} when this
happens: in the Schr\"{o}dinger picture, the state changes continuously,
absent observational input. These are two different ways of keeping track of
the information available to an agent who learns nothing new from observation,
if that information is to continue reliably to predict statistics for outcomes
of future measurements on the system.

By denying the completeness assumption, Wigner's view of the quantum state
removes much of the sting of the measurement problem. On this view there is no
inconsistency between the right hand side of (\ref{2}) and an observation
report of a definite (up or down) position of the apparatus's pointer after it
has interacted with the system. But the view faces difficulties of its own. If
the state vector is only a shorthand expression of the relevant part of
\textit{our} information, then the state vector should be updated whenever
\textit{one} of us learns of the result of a relevant observation, even though
that new information remains unavailable to others.

Wigner was aware of this consequence of his view, even highlighting it in his
"friend" paradox.\cite[pp.179-81]{4} If a friend updates "our" information on
learning the outcome of a quantum measurement on a non-trivially superposed
state conducted in an isolated laboratory, then anyone outside the laboratory
should immediately adopt a new quantum state for the entire laboratory
(including the "friend" as a physical object) that is in principle empirically
distinguishable from the non-trivial superposition resulting from the linear
evolution of its prior quantum state. The decision to count the "friend" as
one of us has empirical consequences.

To avoid the threat of solipsism, Wigner concluded that by becoming
\textit{consciously} aware of the outcome of a quantum measurement, someone
could affect outcomes of \ future observations in a way that could not be
brought about by any interaction with an unconscious physical system. This
view now requires modification of quantum theory, and acknowledgment that Von
Neumann's process 1 is a physical process that supersedes linear evolution
just when the outcome of a quantum measurement is consciously registered. This
is clearly not a satisfactory solution to the measurement problem. It involves
not only modifying quantum theory, but replacing the unmodified theory with a
vaguely formulated mixture of physics and psychology. Wigner's "friend"
paradox is a \textit{reductio ad absurdum} of the view that the state vector
is only a shorthand expression of the relevant part of \textit{our}
information. But a natural variant of the view escapes the paradox and offers
renewed hope of dissolving the measurement problem.

The key is to recognize that differently situated agents, with different
information available to them, should ascribe different quantum states to one
and the same system. This does not make the quantum state subjective. A system
has an objective quantum state relative to the information accessible to an
agent in each specific situation: but in so far as the accessible information
depends on the physical situation of an (actual or potential) agent, so too
does the quantum state for that agent-situation. This point has been
explicitly recognized by some\cite{10},\cite{11},\cite{12},\cite{13}%
,\cite{14}. Brun \textit{et al}.\cite{15} show how much state assignments can
differ on this view of a quantum state as a shorthand expression of that part
of \textit{a situated agent's} information which is relevant for predicting
(as far as possible) the outcomes of measurements on a quantum
system.\footnote{Accepting the relativization of quantum state assignments to
agent-situation also makes it easier to reconcile quantum theory with
violations of Bell inequalities without any physical nonlocality and in
conformity with fundamental Lorentz invariance (see \cite{12}, \cite{13},
\cite{14}, \cite{16},\cite{17}).\ }

On this latest view, a system is not \textit{in} a quantum state, and does not
\textit{have} a (unique) quantum state.\cite{13} Rather, it may
be\ \textit{assigned} a quantum state by an agent for the purposes of applying
quantum theory, and the state to be assigned is a function of the (actual or
hypothetical) physical (and \textit{therefore} epistemic) situation of the one
who assigns it. Relativizing a quantum state assignment to the information
available to an agent helps to dissolve the measurement problem, but it does
not yet answer an important question: How is an agent to update her quantum
state on learning the result of a quantum measurement?

Dirac and Von Neuman took measurement of an observable to leave a system in an
eigenstate of the corresponding operator with eigenvalue equal to the measured
value. This gives a unique updating rule only following the measurement of an
observable with non-degenerate eigenvalues whose eigenvectors span the
system's Hilbert space. L\"{u}ders refined this to the following unique
updating rule for a measurement of $A$ that locates its value in $\Delta$:%
\begin{equation}
\rho\longrightarrow\rho^{\prime}=\frac{P^{A}(\Delta)\rho P^{A}(\Delta
)}{Tr[P^{A}(\Delta)\rho P^{A}(\Delta)]}, \label{Luders rule}%
\end{equation}
where $P^{A}(\Delta)$ projects onto a subspace of vectors, each\ of which
predicts that a measurement of observable $A$ will (with probability 1) find a
value in set $\Delta$. A state updated in accordance with this rule will (with
probability 1) yield the same outcome in an immediately repeated measurement
of $A$, while leaving unchanged the relative probabilities for outcomes in a
measurement of commuting observable $B$ compatible with that outcome for $A$.
Despite figuring prominently in some views of quantum theory, \cite{18}%
,\cite{19},\cite{20} L\"{u}ders' rule is clearly an incorrect way to update a
quantum state after many actual measurements.\cite{3},\cite{21} The momentum
of a neutron will change if it is measured by observing the track of a
(previously stationary) recoil proton in a bubble chamber: measuring the
position of a photon absorbs it at a localized detector. Pauli classified such
measurements as of the second kind in order to \textit{distinguish} them from
measurements conforming to Von Neumann's process 1.

Actual measurements are more realistically modeled by POVMs\cite{22}, and the
state following a POVM $\{E_{i}\}$ may be specified in terms of a set of
measurement operators $\{M_{i}\}$ compatible with (but not defined by) it:%
\begin{equation}
E_{i}=M_{i}^{\dag}M_{i}%
\end{equation}
in which case the appropriate updating rule for a measurement with $i$th
outcome is%
\begin{equation}
\rho\longrightarrow\rho^{\prime}=\frac{M_{i}\rho M_{i}^{\dag}}{Tr[M_{i}\rho
M_{i}^{\dag}]}%
\end{equation}
There is no general rule for associating a particular measurement device with
a POVM, nor for updating the quantum state of a system following a measurement
by that device. One cannot specify either $\left\{  E_{i}\right\}  $ or
$\left\{  M_{i}\right\}  $ theoretically without providing a quantum analysis
of the interaction involved in the measurement. That analysis will depend on
an application of the linear evolution of the combined state of system and
detector (and detector of the detector...), which concludes either by applying
classical physics to some device or with the claim that the "pointer position"
or "click" of some detector was directly observable. The residual measurement
problem is posed by the need to justify this concluding "objectification" step.

\section{The residual measurement problem\label{section 4}}

With advances in technology and increased application of quantum theory to
individual systems (e.g. in quantum optics and quantum computing), its has
become increasingly important to design and characterize quantum measuring
devices. Experimentally, tomography of quantum detectors has been employed to
determine what POVM characterizes a photon detector by "measuring the
measuring device".\cite{23} This involved comparing the detector response (in
the form of a discrete number of "clicks", each intended to indicate detection
of a single photon) with the probability of observed photon number predicted
by applying the Born Rule to a pulse of incident laser light whose quantum
state was taken to be approximately coherent. Increased complexity makes it
harder and harder to model the operation of a detector quantum mechanically as
a linear interaction between system and detector. Braginsky \textit{et
al}.\cite{24} even said

\begin{quote}
The Schr\"{o}dinger equation cannot tell us the connection between the design
of the measuring device and the nature of the measurement, because the
Schr\"{o}dinger equation neither describes nor governs the process of
measurement. (p.38)
\end{quote}

Their reason for saying this was that while the Schr\"{o}dinger equation is
reversible and deterministic, the reduction of the wave-function on
measurement is irreversible and non-deterministic. But as Wigner already
recognized, this does not prevent one from modeling the system/detector
interaction by the linear evolution of their joint state: it merely requires
one to accept that the measurement is not complete until the detector records
a "click" in a way that (randomly) singles out one component of the resulting superposition.

This is the residual measurement problem: Given a superposed entangled state
(such as that of quantum system and quantum detector), under what
circumstances is it legitimate to infer that (at least) one of the entangled
systems \textit{has} some definite property, with probability given by the
Born Rule? The completeness assumption implied that this is never a legitimate
inference: but that assumption has now been rejected along with a
representational view of the quantum state. The no-go theorems of
Gleason\cite{25}, Bell\cite{26}, Kocken and Specker\cite{27} (among others)
imply that this is not always a legitimate inference. To resolve the residual
measurement problem we need to formulate and defend a better answer. This can
be done by applying the quantum theory of decoherence in the right way.

\section{How to resolve it\label{section 5}}

In explaining how quantum decoherence helps solve the residual measurement
problem, it will be useful to be able to refer to a simple model of
decoherence introduced by Zurek\cite{28} and further discussed in Cucchetti,
Paz and Zurek\cite{29}. Consider a single quantum system $A$ interacting with
a second "environment" system $E$. $A$ is a single qubit, and its environment
$E$ is modeled by a collection of $N$ qubits. One can think of each qubit as
realized by a spin
${\frac12}$
system, so that $\left\vert \Uparrow\right\rangle $ ($\left\vert
\Downarrow\right\rangle $) represent $z$-spin up (down) eigenstates of the
Pauli spin operator $\hat{\sigma}_{z}$ of $A$, while $\left\vert
\uparrow\right\rangle _{k}$ ($\left\vert \downarrow\right\rangle _{k}$)
represent $z$-spin up (down) eigenstates of $\hat{\sigma}_{z}^{k}$ for the the
$k$th environment spin subsystem.

The individual Hamiltonians $\hat{H}^{A}$, $\hat{H}^{E}$ of $A$ and $E$ are
assumed to be zero, while the interaction Hamiltonian $\hat{H}^{AE}$ has the
form%
\begin{equation}
\hat{H}^{AE}=%
\frac12
\hat{\sigma}_{z}\otimes\sum_{k=1}^{N}g_{k}\hat{\sigma}_{z}^{k}.
\label{AE interaction}%
\end{equation}
If $A$, $E$ are assumed to be initially assigned pure, uncorrelated states%
\begin{align}
\psi_{A}  &  =\left(  a\left\vert \Uparrow\right\rangle +b\left\vert
\Downarrow\right\rangle \right)  ,\label{initial A state}\\
\psi_{E}  &  =\prod\limits_{k=1}^{N}\left(  \alpha_{k}\left\vert
\uparrow\right\rangle _{k}+\beta_{k}\left\vert \downarrow\right\rangle
_{k}\right)  \label{product form}%
\end{align}
then the initial state%
\begin{equation}
\Psi(0)=\psi_{A}\otimes\psi_{E}%
\end{equation}
evolves according to the Schr\"{o}dinger equation, becoming
\begin{equation}
\Psi(t)=\left(  a\left\vert \Uparrow\right\rangle \left\vert \mathcal{E}%
_{\Uparrow}(t)\right\rangle +b\left\vert \Downarrow\right\rangle \left\vert
\mathcal{E}_{\Downarrow}(t)\right\rangle \right)  \label{AE state at t}%
\end{equation}
at time $t$ where%
\begin{equation}
\left\vert \mathcal{E}_{\Uparrow}(t)\right\rangle =\prod\limits_{k=1}%
^{N}\left(  \alpha_{k}e^{ig_{k}t}\left\vert \uparrow\right\rangle _{k}%
+\beta_{k}e^{-ig_{k}t}\left\vert \downarrow\right\rangle _{k}\right)
=\left\vert \mathcal{E}_{\Downarrow}(-t)\right\rangle .
\label{Environment states at t}%
\end{equation}
The state of $A$, calculated by tracing over the Hilbert space of $E$, is
therefore%
\begin{equation}
\hat{\rho}_{A}(t)=\left\vert a\right\vert ^{2}\left\vert \Uparrow\right\rangle
\left\langle \Uparrow\right\vert +ab^{\ast}r(t)\left\vert \Uparrow
\right\rangle \left\langle \Downarrow\right\vert +a^{\ast}br^{\ast
}(t)\left\vert \Downarrow\right\rangle \left\langle \Uparrow\right\vert
+\left\vert b\right\vert ^{2}\left\vert \Downarrow\right\rangle \left\langle
\Downarrow\right\vert .
\end{equation}
The coefficient $r(t)=\left\langle \mathcal{E}_{\Uparrow}(t)|\mathcal{E}%
_{\Downarrow}(t)\right\rangle $ appearing in the off-diagonal terms of
$\hat{\rho}_{A}$ here is%
\begin{equation}
r(t)=\prod\limits_{k=1}^{N}\left[  \cos2g_{k}t+i\left(  \left\vert \alpha
_{k}\right\vert ^{2}-\left\vert \beta_{k}\right\vert ^{2}\right)  \sin
2g_{k}t\right]  .
\end{equation}
Cucchetti, Paz and Zurek\cite{29} show that $\left\vert r(t)\right\vert $
tends to decrease rapidly with increasing $N$ and very quickly approaches zero
with increasing $t$. More precisely, while $\left\vert r(t)\right\vert ^{2}$
fluctuates, its average magnitude at any time is proportional to $2^{-N}$,
and, for fairly generic values of the $g_{k}$, it decreases with time
according to the Gaussian rule $\left\vert r(t)\right\vert ^{2}\varpropto
e^{-\Gamma^{2}t^{2}}$, where $\Gamma$ depends on the distribution of the
$g_{k}$ as well as the initial state of $E$. This result is relatively
insensitive to the initial state of $E$, which need not be assumed to have the
product form (\ref{product form}), though if the environment is initially in
an eigenstate of (\ref{AE interaction}) $\left\vert r(t)\right\vert =1$ so the
state of $A$ will suffer no decoherence. Since $r(t)$ is an almost periodic
function of $t$ for finite $N$, it will continue to return arbitrarily closely
to 1 at various times: but for $N$ corresponding to a macroscopic environment
Zurek\cite{28} estimated that the corresponding "recurrence" time exceeds the
age of the universe.

Zurek\cite{30} called the unitary evolution (\ref{2}) a
\textit{premeasurement}, and stressed that no actual measurement can be said
to have taken place until $A$ interacts with its environment. This interaction
may be represented by (\ref{AE interaction})\ in an extension of this simple
model. The result is this generalization of (\ref{AE state at t})%
\begin{equation}
\Psi(t)=\left(  a\left\vert +\right\rangle \left\vert \Uparrow\right\rangle
\left\vert \mathcal{E}_{\Uparrow}(t)\right\rangle +b\left\vert -\right\rangle
\left\vert \Downarrow\right\rangle \left\vert \mathcal{E}_{\Downarrow
}(t)\right\rangle \right)  , \label{tridecomposition}%
\end{equation}
where $\left\vert \mathcal{E}_{\Uparrow}(t)\right\rangle ,\left\vert
\mathcal{E}_{\Downarrow}(t)\right\rangle $ are given by
(\ref{Environment states at t}). Zurek\cite{28} defends a slight
generalization of this model to an \textit{n}-dimensional Hilbert space
$\mathcal{H}_{S}$ and \textit{n+1} dimensional Hilbert space $\mathcal{H}_{A}$
as an idealized model of \textit{measurement} on two grounds.

First, the form (\ref{tridecomposition}) preserves the correlation between
component states of $S$,$A$ under the influence of the environmental
interaction (\ref{AE interaction}), and thereby singles out a "pointer basis"
$\left\vert \Uparrow\right\rangle ,\left\vert \Downarrow\right\rangle $ of
states in $\mathcal{H}_{A}$ that correlate with preferred states $\left\vert
+\right\rangle ,\left\vert -\right\rangle $ of $S$. This is important, since
the right-hand side of (\ref{2}), representing the state after the
premeasurement, may be expressed as an entangled superposition in a continuous
infinity of other orthogonal bases of $\mathcal{H}_{A}$. But none
of\textit{\ these} correlations will be preserved under (\ref{AE interaction}%
). This defines the interaction as a measurement of an observable $Q$ on $S$
corresponding to a self-adjoint operator $\hat{Q}$ of which $\left\vert
+\right\rangle ,\left\vert -\right\rangle $ are eigenvectors. Second, as
$\left\vert \mathcal{E}_{\Uparrow}(t)\right\rangle ,\left\vert \mathcal{E}%
_{\Downarrow}(t)\right\rangle $ quickly become (and remain indefinitely) very
nearly orthogonal, the reduced state $\rho_{SA}$ similarly stably becomes very
nearly diagonal in the $\left\{  \left\vert +\right\rangle ,\left\vert
-\right\rangle \right\}  \times\left\{  \left\vert \Uparrow\right\rangle
,\left\vert \Downarrow\right\rangle \right\}  $ basis of $\mathcal{H}_{S}$
$\otimes\mathcal{H}_{A}$. Zurek apparently believes this endows claims about
both the value of $Q$ and the value of the "pointer position" $P$
(corresponding to operator $\hat{P}$ on $\mathcal{H}_{A}$ with eigenvectors
$\left\vert \Uparrow\right\rangle ,\left\vert \Downarrow\right\rangle $) with
enough significance to justify application of the Born Rule, warranting one
confidently to expect these values rapidly and stably to become both definite
and correlated.

The idealizations involved in this simple model are so severe that it can
reasonably be applied to few if any actual cases in which physicists take
themselves to measure the value of a magnitude on a quantum system. It covers
only measurements of the first kind: it assumes the quantum states of $A$ and
$E$ are initially pure, so that the initial quantum states of $S$ and $A$ are
unentangled and both are unentangled with the state of $E$: it further assumes
the initial state of $A$ is $\left\vert \Uparrow\right\rangle _{A}$: it
neglects any effects of\ the individual Hamiltonians of $S$,$A$,$E$ on the
evolution of the total state: and it assumes that the environment does not
interact with $S$ directly, or with $A$ during the interaction between $S$ and
$A$. But if the model is generalized to relax these idealizations, it is not
clear that it retains the features that Zurek took to ground its defense as a
model of \textit{measurement}.

According to Zurek\cite{30}

\begin{quotation}
In the idealized case, the preferred basis was distinguished by its ability to
retain perfect correlations with the system in spite of decoherence. This
remark will serve as a guide in other situations. It will lead to a
criterion---the predictability sieve---used to identify preferred states in
less idealized circumstances. (p.734)
\end{quotation}

However, his applications of the predictability sieve in less idealized
circumstances typically concern only \textit{two} systems: while these are
often labeled $A$ and $E$, a third system $S$ plays no role. So he effectively
abandons his first line of defense that his idealized model is of
\textit{measurement}. But by doing so he suggests a more flexible analysis of
the role of decoherence in measurement.

The idea is to drop any requirement that refers to the post-measurement
condition of the measured system itself, and concentrate only on securing the
direct applicability of the Born Rule to significant claims about the "pointer
position" of a system thought to have interacted with it. As we saw, the
subsequent state of the system depends on details of the measurement
interaction that are irrelevant to justifying application of the Born Rule to
claims about the outcome recorded by a detector. One can think of the measured
system as effectively absorbed into an environment $E=E^{\prime}\oplus S$ that
also includes degrees of freedom of a physical apparatus independent of its
"pointer" degree of freedom. The "pointer position" $P$ is then represented by
an operator $\hat{P}$ in a Hilbert space $\mathcal{H}_{A}$ spanned by a
preferred basis of its eigenstates, the "pointer states". The total state is
then represented in $\mathcal{H}_{A}$ $\otimes$ $\mathcal{H}_{E}$.

So consider an interaction between a system $S$ and an "apparatus" $A\oplus
E^{\prime}$ with initial states $\left\vert \psi_{i}\right\rangle
_{S},\left\vert \psi_{0j}\right\rangle $ respectively, where%
\begin{equation}
\left\vert \psi_{0j}\right\rangle =\left\vert \varphi_{0}\right\rangle
_{A}\otimes\left\vert \varphi_{j}\right\rangle _{E^{\prime}}.
\label{pure apparatus}%
\end{equation}
Here $\left\vert \varphi_{0}\right\rangle _{A}$ represents the "ready to
measure" state of the "pointer" subsystem, and $\left\vert \varphi
_{j}\right\rangle _{E^{\prime}}$ is some basis vector in an environment-system
Hilbert space $E^{\prime}$ that also represents degrees of freedom of the
physical apparatus. To serve as a measurement, the interaction should set up a
correlation between the initial state of the system and the value of a
"pointer position" magnitude $P$\ on the apparatus. In the absence of
interaction with the environment, such an interaction may be thought of as
follows (for some normalized $\left\vert \chi_{ijk}\right\rangle _{S}$ ,
$\left\vert \chi_{ijk}\right\rangle _{E^{\prime}}$)
\begin{equation}
\left\vert \varphi_{0}\right\rangle _{A}\otimes\left\vert \psi_{i}%
\right\rangle _{S}\otimes\left\vert \varphi_{j}\right\rangle _{E^{\prime}%
}\longrightarrow\left\vert \varphi_{i}\right\rangle _{A}\otimes%
{\textstyle\sum\limits_{k}}
\alpha_{ijk}(\left\vert \chi_{ijk}\right\rangle _{S}\otimes\left\vert
\chi_{ijk}\right\rangle _{E^{\prime}}):
\end{equation}
and so, by linearity,%
\begin{equation}
\left\vert \varphi_{0}\right\rangle _{A}\otimes\sum_{i}c_{i}\left\vert
\psi_{i}\right\rangle _{S}\otimes\left\vert \varphi_{j}\right\rangle
_{E^{\prime}}\longrightarrow\sum_{i}c_{i}(\left\vert \varphi_{i}\right\rangle
_{A}\otimes%
{\textstyle\sum\limits_{k}}
\alpha_{ijk}(\left\vert \chi_{ijk}\right\rangle _{S}\otimes\left\vert
\chi_{ijk}\right\rangle _{E^{\prime}})).
\end{equation}
But suppose interaction between $A$ and $E^{\prime}$ rapidly renders
environmental states corresponding to distinct values of $i$ effectively
orthogonal\linebreak($_{E^{\prime}}\langle\chi_{i^{\prime}jk}|\chi
_{ijk}\rangle_{E^{\prime}}\rightarrow0$ for $i^{\prime}\neq i$). Then the
reduced state $\rho_{A}$ will stably approach diagonal form in the $\left\vert
\varphi_{i}\right\rangle _{A}$ basis. A more realistic model would take the
initial "apparatus" state to be a mixture of states of the form
(\ref{pure apparatus}), with variable $j$. The reduced state $\rho_{A}$ will
stably approach diagonal form in the $\left\vert \varphi_{i}\right\rangle
_{A}$ basis in this model also. While this remains a very crude model of
measurement, I think it will help to indicate the role of decoherence in
establishing the applicability of the Born Rule to the results of
measurements, as represented by significant claims about the value of a
magnitude ("the pointer position") on a system that serves as an apparatus.

The residual quantum measurement problem was to answer the question: Given a
superposed entangled state, under what circumstances is it legitimate to infer
that (at least) one of the entangled systems \textit{has} some definite
property, with probability given by the Born Rule? When this question was
posed in the previous section, it may have appeared that the relevant
entangled state is that of a quantum system and quantum detector(s) following
their interaction. Now we see the importance of acknowledging the role of an
environment that makes quantum system plus quantum detector(s) itself an open
system.\footnote{In fact the relevant environment includes subsystems of the
detector(s) corresponding to degrees of freedom other than those associated
with the "pointer observables" whose positions are intended to record an
outcome of the measurement.} In terms of the simplified model, environmental
interactions transfer the relevant entanglement from $S+A$ to $A+E$. Their
effect is then to delocalize the phase of $A$ into $E$ so that the reduced
state of $A$ rapidly and stably becomes very nearly diagonal in the "pointer basis".

According to the pragmatist interpretation I outline in \cite{14},
\textit{this is exactly what is required to license application of the Born
Rule to claims about the position of }$A$\textit{'s\ pointer}. The Born Rule
may be applied only to claims with a well-defined meaning. The significance of
a claim depends on the context, as specified by the nature and extent of
decoherence. When there is extensive and robust decoherence in the "pointer
basis", a claim about the pointer position is richly significant, since it
supports dynamic and measurement inferences: in this context, repeated
measurements of the pointer position do give the same result if repeated
quickly enough, and one can consistently take such results to confirm that the
pointer position evolves continuously, whether or not it is measured. The Born
Rule is not limited in its application to measurement outcomes (however these
may be characterized). But the claims to which it may justifiably be applied
do include some that we endow with the additional significance of reporting
the outcome of some measurement, whether in a laboratory experiment or
elsewhere (e.g. in the context of measurement-based quantum computation).

Quantum theory cannot account for the fact that claims about the values of
magnitudes have truth-values or that measurements have outcomes for the simple
reason that its application \textit{presupposes} that they do---or more
specifically that the set of significant claims of the form $\mathbf{Q\in
\Delta}$ concerning a system to which the Born Rule is applicable contains a
Boolean algebra of events over which a Boolean homomorphism defines a
(classical) truth-value assignment. But it is not at all surprising that a
measurement has some outcome. Quantum theory is empirically based on
observations of measurement outcomes: and application of the Born Rule to a
corresponding claim leads one to assign some definite (though sometimes small)
credence to whatever outcome is observed. Since the application of the Born
Rule is not limited to the outcomes of measurements, it is equally
unsurprising that the world can be truly described by many other
claims---enough to constitute a rich description in non-quantum terms.

Quantum decoherence through environmental delocalization of phase has often
been take to explain the appearance of a classical realm, though both the
meaning and truth of this claim remain controversial. One could take the claim
to be that classical physics is reducible to quantum theory, or (in a
different usage) that quantum theory reduces to classical physics in some
appropriate limit (e.g. as $\hbar\rightarrow0$). The concepts of reduction
involved here cry out for further clarification. But the pragmatist
interpretation of quantum theory outlined in \cite{14} encourages no such
reductionist claim, since on this interpretation one does not use quantum
states to describe a "quantum world" within which a (quasi)classical realm may
emerge. An agent is well-advised to use quantum states as a guide in assessing
the content and credibility of various claims about physical systems, and
sometimes is justified in wholly believing some such claims. But the truth of
those claims does not then rest on that of any underlying \textit{quantum}
truths about these or any other systems. If that is what Bohr meant by his
reported remark "There is no quantum world" then he was correct.\cite{31}

Nevertheless, an agent applying quantum theory may be warranted in making
claims that would also be warranted within classical physics, and these claims
may be justified by suitable application of the quantum theory of
environmental decoherence. Hornberger, Sipe and Arndt\cite{32} applied the
quantum theory of environmental decoherence to show why one would be warranted
in expecting the visibility of the interference pattern to decrease in the
experiment of Hackerm\"{u}ller \textit{et} \textit{al}.\cite{33} as they
increased the laser heating of the fullerene molecules that passed through
their interferometer until the pattern accorded with classical physics. The
theory of quantum Brownian motion\cite{34} already provides a simplified model
of how claims about the classical trajectory of a quantum particle bound in a
harmonic potential may be warranted as a result of environmental decoherence.
A more sophisticated application of the quantum theory of decoherence would be
expected to provide an even stronger warrant for claims about the fixed
locations of the fullerenes deposited on the silicon surface in the experiment
of Juffman \textit{et al}.\cite{35}---claims that would also be a consequence
of applying classical physics to their initial locations on the surface.

There are cases in which quantum theory itself seems to imply a claim about
the value of a magnitude. Ehrenfest's theorem is even adduced as a case in
which a law of classical mechanics (Newton's second law of motion) follows
from a law of quantum mechanics in suitable circumstances. For a single
particle subject to a potential $\Phi(\mathbf{r},t)$ the following is an easy
consequence of the Schr\"{o}dinger equation:
\begin{equation}
\frac{d}{dt}\left\langle \mathbf{p}\right\rangle =-\left\langle \mathbf{\nabla
}\Phi\right\rangle \label{Ehrenfest}%
\end{equation}
This closely resembles Newton's second law in the form $d\mathbf{p}%
/dt=-\mathbf{\nabla}\Phi$. But what it actually says is rather different: that
the rate of change of the \textit{expectation value} of momentum equals minus
\textit{the expectation value} of the gradient of the potential. A claim about
the expectation value of a magnitude is not a claim about the actual value of
that magnitude. According to the pragmatist interpretation outlined in
\cite{14} its function is not to describe values of magnitudes but rather to
guide an agent's \textit{expectation} as to their values. Indeed, the
terminology of "expectation values" nicely captures this function.
Mathematically, an expected value is the mean of a random variable in a
probability distribution---in this case calculated by application of the Born
Rule to a quantum state. But how can such applications be justified in the
case of (\ref{Ehrenfest}) when the magnitudes on either side of the equation
are represented by non-commuting operators that have no joint distribution in
any quantum state?

The quantum theory of decoherence can help answer this question. If the
particle in question is subject to the right kind of environmental interaction
(e.g. weakly coupled to an oscillator bath, as in the case of quantum Brownian
motion\cite{34}) its reduced quantum state will rapidly and robustly become
approximately diagonal in a basis of coherent states. This will justify
application of the Born Rule both to claims about its position $\mathbf{r}$
(and functions such as $\Phi(\mathbf{r})$) and to claims about its momentum
$\mathbf{p}$. When the Wigner function of this quantum state is non-negative
everywhere it is even possible to represent these as deriving from a
corresponding joint probability distribution, though such representation is
merely a technical convenience. Any agent applying quantum theory is then
warranted in believing both that the particle has some reasonably well defined
position and momentum and that these will almost certainly evolve continuously
in accordance with Newton's second law. If the agent is warranted in ascribing
a \textit{particular} narrow Gaussian wave-function as the initial quantum
state, then the warrant extends to confident belief in claims restricting the
particle's position and momentum to values close to the peak of that evolving Gaussian.

Among many similar examples in which the quantum theory of decoherence
justifies replacing quantum expectation values by corresponding claims about
non-quantum magnitudes is this important claim:

\begin{quote}
When a constant voltage $V$ is applied across a Josephson junction, an
alternating current $I$ with frequency $2(e/h)V$ flows across the junction.
\end{quote}

Each of the voltage $V$ and the current $I$ here results from identification
of a non-quantum magnitude with an expectation value in a quantum analysis
that associates "macroscopic wave-functions" with the superconductor on either
side of the junction. In any experimental realization of this AC\ Josephson
effect, it is environmental decoherence acting on these wave-functions that
justifies application of the Born Rule, though no simple model of such
decoherence may be available.

\section{Measuring photons\label{section 6}}

Measurement of photons has always been difficult to reconcile with the
Dirac-Von Neumann account of measurement.\cite{36} This process can be
accommodated within the more general POVM framework by associating it with
measurement operators such as $\left\{  M_{n}\right\}  $ $(n=0,1,...)$, each
element of which projects a Fock state of the quantized electromagnetic field
onto the vacuum state:
\begin{equation}
M_{n}=\left\vert 0\right\rangle \left\langle n\right\vert . \label{photons}%
\end{equation}
But notice that these do not yield any subsequent state of the photon, but the
"zero-photon" state of the electromagnetic field---in other words, the
measurement destroys the measured photon. A more general treatment of
measurement of the quantized electromagnetic field may be given in terms of
measurement operators that modify its state in ways that do not correspond
naturally to measurements of photons, such as the coherent state measurement
operators%
\begin{equation}
M_{\alpha}=1/\sqrt{\pi}\left\vert \alpha\right\rangle \left\langle
\alpha\right\vert . \label{coherent states}%
\end{equation}
Nevertheless, many experiments in quantum optics are taken to involve
measurements on single photons, and the Born Rule is applied to claims about
their properties, including even their positions. How can this be justified?

Sinha \textit{et al}.\cite{37} set out to test an important implication of the
Born Rule: In the interference pattern resulting from more than two paths, the
interference terms are the sum of the interference terms in the patterns
resulting from these paths taken two at a time. They matched observed
interference patterns against an instance of this implication in a variety of
experiments involving the interference of light at up to three slits. One of
these involved single photons, detected in coincidence with a second
\textquotedblleft herald\textquotedblright\ photon from an entangled pair by
two avalanche photodiodes. The experimental interference pattern is generated
by moving a multimode optical fiber uniformly across a plane intercepting
light from the slits and counting the relative number of photons detected in
each small region.

One can understand this experiment as a test of the Born Rule only if that
rule is applicable to the experimental data. Perhaps the simplest way to apply
it would be to claims of the form:

\begin{quote}
(X) The position of the photon lies between $x$ and $x+\Delta x$
\end{quote}

Here $\Delta x$ represents a small interval of positions in a direction
perpendicular to the slits and the optical axis, and the claim concerns the
position in the tracking plane where a photon enters the fiber. But there are
reasons to question the significance of a claim of this form here. There is no
well-behaved position operator in relativistic quantum theories, and it is
difficult if not impossible to understand a relativistic quantum field theory
as \textit{describing} localized particles.\cite{38},\cite{39},\cite{40} Any
talk of photons acquires whatever theoretical significance it has from
applications of the quantum theory of the electromagnetic field. In cavity
quantum electrodynamics, for example, it is often convenient to use the term
'photon' when considering quantized energy states of the field in the cavity.

Photon talk also acquires practical significance in quantum optics experiments
like that of Sinha \textit{et al}. In the course of their review of several
such experiments, Zeilinger \textit{et\ al}.\cite{41} say

\begin{quote}
...the quantum state is simply a tool to calculate probabilities.
Probabilities of the photon being somewhere? No, we should be even more
cautious and only talk about probabilities of a photon detector firing if it
is placed somewhere. One might be tempted, as was Einstein, to consider the
photon being localized at some place with us just not knowing that place. But,
whenever we talk about a particle, or more specifically a photon, we should
only mean that which a 'click in the detector' refers to.
\end{quote}

Indeed, Sinha \textit{et al}. take the Born Rule to specify the probability
(density) \textit{to find} or \textit{detect} a particle at position $r$, not
for a particle to \textit{be} at $r$. But of course such appeals to
experimental practice do not answer the interpretative question as to exactly
when and why an agent applying quantum theory is entitled to claim that a
particle has been detected at position $r$. In the pragmatist interpretation
outlined in \cite{14} the quantum theory of decoherence can help answer this question.

In the present situation, no significant decoherence occurs before the
quantized electromagnetic field interacts through the photoelectric effect
with electrons in the avalanche photodiodes. Such interaction directly endows
a claim about a system with a rich empirical significance only at the
photodiodes themselves. But it may be taken thereby indirectly to render
significant some claim about photon position at the interception plane across
which the multimode optical fiber is tracked. For the \textquotedblleft
backtracking\textquotedblright\ inference from a warranted claim about the
value of a magnitude at the photodiode to some claim about photon position in
the interception plane may seem justified here by the assumption that the
multimode fiber provided the only available channel through which the field
could propagate. This suggests that such an inference could legitimize
application of the Born Rule to claims of the form (X) concerning the position
of photons in the interception plane. But that is not quite right.

The photodiode is designed to produce a substantial (milliampere) electron
current when light injects even a single electron into the depleted region by
the photoelectric effect. In practical terms, the subsequent amplification
process in the photodiode is highly irreversible, but it is not this
irreversibility but rather the associated decoherence that ensures that some
claim about a magnitude on a system within the diode directly acquires
substantial empirical significance. This is not, however, a claim about the
position of a photon.

Neither I\ nor the authors of Sinha \textit{et al}. have presented any quantum
model of decoherence at the avalanche photodiode in the experiment described.
But it is not necessary to advance such a model to be sure that almost
immediately after any electron is freed by the photoelectric effect a claim
about \textit{some} system within the diode will acquire substantial empirical
significance through interactions with the system's environment. Exactly what
this system is and what magnitude figures in the claim is immaterial to the
operation of the photodiode. So in practice the effective nature and location
of the 'detector click' cannot be specified more precisely than somewhere
within the photodiode's "window" of sensitivity. The spatial window is
comparable to the 65 micron core of the multimode fiber whose aperture probes
the interference: the temporal window is of the order of nanoseconds. But the
practical impossibility of verifying, or even unambiguously specifying, any
more precise claim about a system at the photodiode here is not what
determines the limits of empirical significance of such magnitude claims and
whether this justifies applying the Born Rule to them.

Electrons ejected from a metal surface by the photoelectric effect have a
definite energy, as experiments reveal. Similarly, it is generally assumed
that when an electron is ejected into the depletion region of a photodiode it
has a definite energy, equal to the difference between the energy of the
photon involved and the electron's binding energy. This goes along with the
idea that the electron was ejected from a definite energy level---an
assumption that implicitly depends on a model of decoherence for the
crystalline structure of the semiconductor. So a claim of the form

(Y) The kinetic energy of an ejected electron is $E>0$\newline owes its
substantial empirical significance to energy decoherence affecting systems
within the photodiode. One may take this as a claim about an event
corresponding to detection of the photon---to what Zeilinger \textit{et\ al}.
call a click in the detector---even prior to the electron avalanche it
induces. On the interpretation sketched in \cite{14}, it is claims such as
this that indirectly justify application of the Born Rule to claims about
photon position in the interception plane.

Taken literally, the claim (X) has no empirical content in this experiment
since it cannot be grounded either in the quantum theory of light or in
experimental practice. To justify application of the Born Rule the claim must
therefore be reformulated or reinterpreted. So consider the following
reformulation, which, as we have seen, accords well with how many physicists
express themselves.

\begin{quote}
(X$^{\prime}$) The position of the photon is detected between $x$ and
$x+\Delta x$
\end{quote}

The content of (X$^{\prime}$) is quite unclear as it stands. But one can now
use an inferentialist account of content to clarify it. The key is to link the
content of (X$^{\prime}$) to that of (Y) by taking it as an essential part of
the content of (X$^{\prime}$) that it follow by a justified (though not
deductively valid) inference from (Y). This is not to equate the contents of
(X$^{\prime}$) and (Y)---these claims certainly don't mean the same thing. But
(X$^{\prime}$) owes its substantial empirical significance to this close
inferential relation to (Y), and that is what justifies application of the
Born Rule---not to (X) but to (X$^{\prime}$). Here and elsewhere, the
empirical significance of a magnitude claim about a system (in this case, the
claim (X$^{\prime}$)) hinges on what actually happens \textit{later}.
Moreover, the pivotal later event may involve a distinct system (the ejected
electron, in this case).

\section{Measurement and Quantum Fields\label{section 7}}

We saw that measurement of photons is just a special case of measurement of
the state of the quantized electromagnetic field. This is only one of many
relativistic quantum fields that figure in contemporary physics. That
electromagnetism has well understood manifestations as a classical field even
at low energies makes it a useful example to introduce a discussion of the
special issues quantum field theories raise for the measurement problem.

Traditionally, the outcome of a quantum measurement was taken as a record of
the value of an observable on the measured system. But the POVM\ framework
allows an understanding of the measurement process as giving probabilistic
information about the quantum state of the measured system that may take a
more general form, not necessarily focused on a specific observable.
Measurements represented by the measurement operators (\ref{photons}) and
(\ref{coherent states}) yield different probabilistic information about the
state of a quantized electromagnetic field while (in general) altering that
state: only for the first of these can measurement serve to record the value
of an observable. The information provided by a general quantum measurement is
useful not because it tells one what value some observable had on the measured
system, nor because it tells one what value some observable has acquired as a
result of the measurement, but because it guides expectations concerning the
possible results of future measurements either on the original state or on the
state after this initial measurement.

Freed from the need to think of measurement as a process directed toward
recording the value of some magnitude on a system, an analysis of measurement
on quantum field systems should address two separate questions:\linebreak1)
How can a measurement of a quantum field lead to an outcome?\linebreak2)
\ What makes a claim about a quantum field magnitude significant?

Del\'{e}glise \textit{et al}.\cite{42} report studies of many non-classical
states of the electromagnetic field in a cavity. The field is probed by
Rydberg atoms passed through the cavity whose energy states become entangled
with the state of the cavity and are subsequently observed. Such an atom
functions as a quantum "probe", and the outcome of the field measurement is
objectified by observing the state of atomic excitation. That observation is
performed by ionizing the atom and detecting the emitted electron, a process
that is selective since different energy states are ionized in different
electric field strengths in the detector. Unlike many measurements on the
field, observation of the atom is thought of as a (destructive) measurement of
an observable---its energy. Prior to this measurement, the atom's state was
(typically) entangled, but only with that of the cavity field, so a claim
about its energy lacked significance. But after detection of its ionized
electron a claim about the state of the detector acquires a rich significance
through environmental decoherence of the detector's state. It is only at this
stage that an objective outcome of the cavity field state measurement emerges.
This involves no significant claim about a quantum field magnitude, and no
significant claim about the value of any magnitude on the atom---not even
about its energy. A measurement of the quantum field leads to an outcome only
through decoherence in the atom detector. What is measured is not a quantum
field magnitude.

This illustrates an important generic feature of measurements on a quantum
field system that goes a long way toward answering question (1). The outcome
of such a measurement is recorded not in a claim about the value of a
magnitude on that quantum field, but in a claim about the value of a magnitude
in a detector, even when that outcome is called the result of a measurement of
the field. The state of the field becomes entangled with that of some
subsystem of the detector, which is then decohered by interaction with its
environment. It is this decoherence that gives rise to an outcome of the
measurement---not by "collapsing the state" (instead, the system-detector
entanglement is extended to the environment) but by rendering significant a
claim about the value of a magnitude on the detector. The detector is designed
so that the truth of this claim may be readily checked and/or recorded by
directly examining it.

While the result of the measurement is often expressed in language that
apparently commits its author to the existence and properties of physical
systems associated with a quantum field (as in the example of single photon
detection) such talk should be seen as inferentially grounded in claims about
properties of some distinct probe or detector system involved in the
measurement. This "inferential buck-passing" is not a feature that
distinguishes measurement of quantum fields from measurement of other quantum
systems, as was illustrated by the measurement of the energy of the Rydberg
atoms used to probe the cavity field. In no such case does the objectification
of the measurement outcome require the attribution of the measured property to
the measured system itself, either before or after the measurement.

It is in answering question (2) that a distinctive feature of quantum fields
appears. Along with quantum mechanics and other variants of quantum theory,
quantum field theories share a common framework of quantum states and
probabilities, whose joint function is to advise an agent applying the theory
on the content and credibility of magnitude claims. The Born Rule is
applicable directly only to canonical magnitude claims---those of the form
$\mathbf{\sigma}$ \textbf{has }$\mathbf{Q\in\Delta}$, for $\Delta$ a Borel set
of real numbers, where $Q$ is a magnitude corresponding to self-adjoint
operator $\hat{Q}$ on a Hilbert space (or in some algebra of operators, in a
more abstract form of quantum theory). In a quantum field theory, such
magnitudes include components of fields such as the electric field
$\mathbf{E}(\mathbf{r})$ and magnitudes corresponding to the fermion number
density $\hat{\psi}^{\dag}(\mathbf{r})\hat{\psi}(\mathbf{r})$, but not to the
fermion field $\hat{\psi}(\mathbf{r})$ itself. Fully to specify the content of
a canonical claim one must say what the system $\sigma$\ is.

Call this system $\sigma$\ the \textit{target} of an application of quantum
theory. While targets will vary from application to application, most if not
all systems figuring in canonical claims may be naturally grouped into two
kinds: fields and particles. Besides advising on the content of a predicate
$\mathbf{Q\in\Delta}$ , a quantum state must also advise a user on the content
of the singular or general term $\sigma$ that picks out the target system (or
systems) to which the predicate is applied.

Quantum mechanics is targeted on systems of particles---individually or
collectively. This is true of Dirac's relativistic theory of electrons as well
as non-relativistic quantum mechanics as applied to condensed matter,
molecules, atoms or atomic constituents at low energies. While quantum field
theory is superficially (and mathematically) a theory of fields, it is not
targeted on physical \textit{quantum} field systems. The canonical claims on
which the quantum state of a quantum field theory directly offers advice may
concern either \textit{classical} fields or particles, depending on the
circumstances in which the theory is applied. Consider first the case of the
quantized electromagnetic field.

Kiefer\cite{43},\cite{44} analyzes the interaction of a quantized
electromagnetic field with a quantized scalar matter field ("scalar quantum
electrodynamics"). He describes circumstances in which this interaction has
the effect of decohering the quantum state of the electromagnetic field (in a
Schr\"{o}dinger functional representation) so that the reduced state (after
tracing over the matter field Hilbert space) rapidly becomes diagonal in a
basis of WKB states that approximate a state of the classical electromagnetic
field---with quite well-defined electric field and magnetic vector potential.

Anglin and Zurek\cite{45} give a model in which the state of a quantized
electromagnetic field is decohered by a set of harmonic oscillators
representing a homogenous, linear, dielectric medium. In this model the
reduced state of the field quickly becomes approximately diagonal in a basis
of coherent states. This shows how, in suitable circumstances, environmental
decoherence can select a basis of preferred states that closely approximate a
classical electromagnetic field. According to \cite{14} this is exactly what
is required to endow a claim about the strength of the electric or magnetic
field at a point with the significance required to justify application of the
Born Rule to assign a probability to that claim. While their model
incorporates a number of idealizations that do not hold for an electromagnetic
field in a medium such as the atmosphere, they argue that this result is
robust enough to hold also for electromagnetic radiation at wavelengths down
to the ultraviolet propagating through the air. Moreover the dynamics of the
decohered states will conform closely to classical electromagnetic theory. In
a situation in which an agent is warranted in ascribing a particular coherent
initial quantum state to the quantized electromagnetic field, the Born Rule
will entitle him to expect the electric and magnetic fields to evolve in a way
consistent with classical electromagnetic theory.

Anglin and Zurek also discuss the status of photons and other "particles" in
the light of their model. Referring to the coherent states that provide a
stable, overcomplete basis for the decohered reduced state of a quantum field
as pointer states, they conclude

\begin{quotation}
Each pointer state of a quantum field is surrounded, in Hilbert space, by a
quantum halo --- a set of states which are negligibly decohered from the
pointer state over whatever time period is of interest. When the environmental
noise is weak enough that it does not significantly degrade the pointer states
themselves, this quantum halo is large enough to contain at least a few
particles, excited above the background classical field configuration
represented by the pointer state. We have thus recovered the familiar
field-theoretic dichotomy between background classical fields and
\textit{N}-particle excitations. The relative immunity of the particle
excitations to decoherence, in comparison with the strong decoherence of
superpositions of distinct pointer states, explains the coexistence of
effective classical electrodynamics and coherent propagation of photons. The
\textit{n}-particle excitations are not localized by our homogeneous
environment. All localization occurs in the space of coherent state
amplitudes, and not in position space.\cite[p.7334]{45}
\end{quotation}

This sheds further light on the extent to which application of the quantum
theory of the electromagnetic field endows talk of photons with empirical
content. Even experimenters in quantum optics violate the injunction:
"whenever we talk about a particle, or more specifically a photon, we should
only mean that which a 'click in the detector' refers to". They do so whenever
they speak of \textit{propagation} of photons---through an optical fiber, the
atmosphere, or empty space. A \textit{classical} electromagnetic field
corresponding to a coherent state \textit{can} significantly be said to
propagate. When the coherent state is surrounded by a quantum halo---a space
of "nearby" quantum states---it is tempting to say these states describe
photons propagating with the classical field to which the coherent state
corresponds, even though no quantum state describes any physical system. But
such talk can mislead. One must remember that not every state in the quantum
halo of a coherent state can be thought to contain any definite number of
photons: the halo contains superpositions and mixtures corresponding to
different "photon numbers", that are not decohered by the environment.
Moreover, in an experiment such as that of Ursin \textit{et} \textit{al}%
.\cite{46} if one speaks of creating a pair of entangled photons $a$,$b$ one
of which ($a$) traveled to one detector while the other ($b$) traveled to
another distant detector, this must either be understood as merely a
metaphorical gloss on claims about propagating classical fields, non-linear
sources and "detector clicks", or (consistent with an inferentialist account
of what gives a claim meaning) allowed to stand on its own \textit{provided}
the inferences it is taken to support are carefully circumscribed.

In quantum theory one models target systems by assigning quantum states to
systems and applying the Born Rule. Models in quantum mechanics (relativistic
as well as non-relativistic) can be taken to assign quantum states to the
target systems themselves, be they electrons, atoms or other
particles.\footnote{But this is not necessary. Quantum models often represent
quantum states e.g. of abstract harmonic oscillator or spin systems in an
appropriate Hilbert space model in order to abstract from messy details of
actual target systems.} So the general practice of applying a significant
predicate $\mathbf{Q\in\Delta}$ to electrons, atoms and other particles needs
no defense in quantum mechanics. But a model of a quantum field theory such as
quantum electrodynamics assigns quantum states not to electrons or other
target systems, but to an abstract quantum field system (such as interacting
quantized electromagnetic and charged lepton fields): model systems can no
longer be identified with target systems. So the attribution of location,
momentum, spin, etc. to electrons, atoms and other particles does require
justification in an application of a quantum field theory model. It is clear
that this cannot take the same form as that just given for photon talk in the
context of the quantum theory of electromagnetism. We ascribe energy,
momentum, spin, etc. to elementary particles when applying the quantum field
theories of the Standard Model and think of them as located in a particle
accelerator or in one of the associated detectors used to test the Model. But
unlike the electromagnetic field, neither the fermionic nor the massive
bosonic fields of the Standard Model have manifestations as classical fields.

Anglin and Zurek discuss this issue in their conclusion, where they argue that
the key difference between an application of a quantum field theory where
classical field-like behavior becomes manifest and an application where
classical particle-like behavior becomes manifest is the different character
of environmental coupling present in the two cases. Specifically, while the
electromagnetic field typically couples to its environment through a linear
coupling, the electron field (say) typically couples to its environment
\textit{bi}linearly. They suggest (but do not give a comparable analysis to
prove) that such a coupling will select a basis of \textit{n}-particle states,
rather than coherent states, as those that remain stable under environmental
interactions. According to the interpretation outlined in \cite{14} this is
exactly what is required to endow at least some claims about the values of
magnitudes on the particles present in such a state with\textit{\ }the
significance required to justify application of the Born Rule to assign a
probability to those claims.

This cannot be the whole story. Consider the suggestion that the significance
of claims attributing classical field values requires a linear coupling to a
decohering environment to select a basis of coherent states of a quantized
field, while a bilinear coupling selects out \textit{n-}particle states. This
may help explain the difference between how we talk about certain gases as
collections of particles at ordinary temperatures but as coherent fields in
BECs at extremely low temperatures. But it seems unlikely that decoherence can
ground familiar claims about most of the extremely short-lived "elementary
particles" detected in high energy accelerators, whether bosons or\ fermions.
Perhaps claims apparently ascribing properties to these "particles" could be
rephrased as (non-descriptive) claims about states of their associated quantum
fields, just as talk of $N$ photons in a cavity is often rephrased as talk of
an $N$-photon state. Alternatively, claims about properties of "elementary
particles" could derive their significance from their inferential relations to
significant descriptive claims about outcomes of measurements on their
associated quantum fields, along the lines of claims about positions of photons.

\section{Conclusion\label{section 8}}

One can reconcile the observed outcomes of measurements with quantum theory by
recognizing the non-descriptive role of the quantum state. Decoherence grounds
the significance of claims about these outcomes and their probabilities. But
quantum theory assumes and cannot account for the fact that measurements have
outcomes (still less the \textit{particular} outcome we observe) since it is
not the role of a quantum state to represent a measurement outcome.

\begin{acknowledgement}
This publication was made possible through the support of a grant from the
John Templeton Foundation. The opinions expressed in this publication are
those of the author and do not necessarily reflect the views of the John
Templeton Foundation.
\end{acknowledgement}

\end{document}